\documentclass{article}

\usepackage{geometry}
\usepackage[labelfont=bf]{caption}
\usepackage{graphicx}
\usepackage{setspace}
\usepackage{amsmath,amsthm,bm}
\usepackage{algorithm}
\usepackage{tabls}
\RequirePackage{hyperref}
\usepackage[utf8]{inputenc} 

\makeatletter
\def\@listiii{\leftmargin\leftmarginiii
              \labelwidth\leftmarginiii
              \advance\labelwidth-\labelsep
              \topsep\z@
              \parsep\z@
              \partopsep\z@
              \itemsep\topsep}
\makeatother

\newcommand{\Score}{\mathit{Score}}
\newcommand{\Prob}{\operatorname{P}}

\newcommand{\given}{\operatorname{|}}
\newcommand{\X}{\mathbf{X}}

\newcommand{\D}{\mathcal{D}}
\newcommand{\G}{\mathcal{G}}
\newcommand{\M}{\mathcal{M}}
\newcommand{\LL}{\mathcal{L}}

\newcommand{\PXi}{\Pi_{X_i}}
\newcommand{\XPi}{X_i \given \PXi}
\newcommand{\T}{\bm{\Theta}_{X_i}}

\newcommand{\Gmax}{\G_{\mathit{max}}}
\renewcommand{\S}{\mathbf{S}}
\newcommand{\Smax}{S_{\mathit{max}}}

\newcommand{\BIC}{\mathrm{BIC}}
\newcommand{\DAG}{\mathrm{DAG}}
\newcommand{\BICg}{\BIC_{\gamma}}

\newgeometry{vmargin={25mm,20mm}, hmargin={20mm,20mm}} 

\title {Learning complex dependency structure of gene regulatory networks from high dimensional micro-array data with Gaussian Bayesian networks}

\author{Catharina E. Graafland$^{1}$ \& Jos\'e M. Guti\'errez$^1$}

\date{%
	$^1$Instituto de F{\'\i}sica de Cantabria, 
	CSIC--Universidad de Cantabria, Avenida de Los Castros, 
	E-39005 Santander, Spain\\[2ex]
	\today
}

\begin{document}
\maketitle

\begin{abstract} 
\noindent Reconstruction of Gene Regulatory Networks (GRNs) of gene expression data with Probabilistic Network Models (PNMs) is an open problem. Gene expression datasets consist of thousand of genes with relatively small samplesizes (i.e. are large-$p$-small-$n$). Moreover, dependencies of various orders co-exist in the datasets. On the one hand Transcription Factor encoding genes (TFs) act like hubs and regulate target genes, on the other hand target genes show local dependencies. In the field of Undirected Network Models (UNMs) -- a subclass of PNMs-- The Glasso algorithm has been proposed to deal with high dimensional micro-array datasets forcing sparsity.
To overcome the problem of complex interaction structure modifications of the default Glasso algorithm are developed that integrate beforehand expected dependency structure in UNMs. In this work we advocate the use of a simple score-based Hill Climbing algorithm (HC) that learns Gaussian Bayesian Networks (BNs) leaning on Directed Acyclic Graphs (DAGs). We compare HC with Glasso and its modifications in the UNM framework on their capability to reconstruct GRNs from micro-array data belonging to the Escherichia Coli genome. We benefit from the analytical properties of the Joint Probability Density (JPD) function on which both directed and undirected PNMs build to convert DAGs to UNMs.
We conclude that dependencies in complex data are learned best by the HC algorithm, presenting them most accurately and efficiently, simultaneously modelling strong local and weaker but significant global connections coexisting in the gene expression dataset. The HC algorithm adapts intrinsically to the complex dependency structure of the dataset, without forcing a specific structure in advance.  
\end{abstract}



\section*{Introduction}

The reconstruction of Gene Regulatory Networks (GRNs) of gene expression data is an open problem and has attracted great deal of interest for decades. GRNs  model interaction structure of genes in a network and are important to understand the biological functions in living organisms as well as the regulation of diseases in them. The increasing availability and improved systematic storage of gene expression data obtained with DNA micro-arrays \cite{faith_many_2008} revolutionized the incorporation of mathematical and computational models to model GRNs in the past two decades\cite{mccall_estimation_2013,delgado_computational_2019}.

Mathematical models range from more complex models relying on sets of differential equations that directly describe dynamic changes in GRNs \cite{de_jong_modeling_2002,chen_modeling_1999} to  simpler models that describe GRNs building on a graph presentation (graphical models or network models). The latter attract attention due to their capacity of visualization of the complex interaction structure of micro-array data in a graph. 
The most simple and widely used example of a graphical model are pairwise Correlation Networks (CNs) \cite{zhang_general_2005}, but lately more advanced Probabilistic Network Models (PNMs) that make use of machine learning algorithms to comprehensively model conditional and/or partial dependencies have gained in popularity.

One of firsts applications of PNMs to reconstruct GRNs was made by Friedman et al.\cite{friedman_using_2000}. Their work substantially promoted later research in the field. They analyzed the use of Bayesian Networks (BNs)-- a subclass of PNMs relying on conditional dependencies modeled in a Directed Acyclic Graph (DAG)-- on data of S.cerevisiae cell-cycle measurements. In their work they set the field to model GRNs with two approaches, multinomial BNs and Gausian BNs. The former relies on discrete data and thus requires continuous micro-array to be discretized, the latter can directly handle continuous micro-array data assuming an overall Gaussian distribution function. Later use of BNs in the field, however, mainly focused on adaption and improving algorithms for multinomial BNs \cite{friedman_using_2000, xing_improved_2017,hartemink_using_2001,peer_inferring_2001, zou_new_2005}, while the application of Gaussian BNs have been left untreated (with some exceptions, e.g.  Wehrli et al. \cite{werhli_comparative_2006} in which Gaussian Bayesian networks applied to GRNs appear in a comparison study).

The Gaussian case though has been widely investigated in the subclass of PNMs that relies on undirected graphs, i.e. Undirected Probabilistic Network Models (UNMs) (also known as undirected Markov random fields or pairwise Markov networks)\cite{dobra_sparse_2004}. 
The most outstanding algorithm of the past decade to learn Gaussian UNMs had its first application on cell-signaling data from proteomics \cite{friedman_sparse_2008}. This algorithm is called the Graphical lasso (Glasso) and estimates the inverse covariance matrix, also concentration or precision matrix, of the Gaussian distribution function. It is successfully applied in the high dimensional setting of gene interaction networks \cite{li_gene_2015} \cite{zhao_cancer_2019}, clustering of networks in bioinformatics \cite{mukherjee_network_2011}, but also in psychology networks \cite{epskamp_tutorial_2018}, risk management\cite{perederiy_bankruptcy_2009}
\cite{chan-lau_lasso_2017} and climate  \cite{zerenner_gaussian_2014}.

The application of Gaussian UNMs on continuous micro-array data have faced two main challenges. First of all, the large-$p$-small-$n$ character of the data. Experiments with micro-arrays contain expression levels of thousands of genes at the same time, but have relative small samplesize. Glasso deals with high dimensional data imposing sparsity in the precision matrix -- and thus in the undirected graph. 
Secondly, degrees of interaction of genes in gene networks are not uniform, instead they are of higher-order. In GRNs Transcription Factors (TFs) regulate the expression of many target genes. However, only relatively few genes encode TFs, while most genes are just `being regulated' by TFs. In a GRN, TF encoding genes can be seen as hub genes that procure outliers in the overall degree distribution \cite{he_understanding_2016}\cite{akesson_comhub_2021} and earlier research on the degree distribution of a GRN suggested characteristics of scale-free networks \cite{barabasi_emergence_1999}.

In recent years quit a lot of modifications of the Glasso algorithm distribution were proposed to better model the complex interaction structure in micro-array data (and other real world datasets). Two of them were especially developed for gene regulatory networks expected to be scale-free \cite{liu_learning_2011} or to consist of hubs \cite{mcgillivray_estimating_2020}. We will refer to them as the Scale-Free Glasso (SFGlasso) and the Hub Glasso (HGlasso). Both modifications of the Glasso algorithm force beforehand expected structure in the estimation of the precision matrix and where shown to outperform Glasso on simulated data that had these characteristics. 



As an alternative in this paper, we build on the initial work of Friedman et al. \cite{friedman_using_2000} and propose the use of Gaussian Bayesian Networks (GBNs) to tackle the complex structure problem in GRNs. Our motivation comes from a recent work showing the strength and simplicity of GBNs for modelling the complex (hub) interaction structure that occurs in high dimensional data, reveling the underlying probabilistic backbone \cite{graafland_probabilistic_2020}.
We here revisit and use the score-based Hill climbing algorithm (HC) that we earlier found to learn best GBNs for high-dimensional complex data \cite{scutari_who_2019}\cite{graafland_probabilistic_2020} and that doesn't assume specific structure in advance. We compare HC with the default approach for learning sparse UNMs, the Graphical lasso (Glasso) and with its modifications SFGlasso and HGlasso.
We do so by reconstructing (the most up to date version of) the dataset used in Yu et al. \cite{yu_enhanced_2017} from the Escherichia Coli (E.Coli) genome, which is entitled to be the best known/encoded organism on earth and from which information on TFs is well documented. 
This simultaneous study and intercomparison of direct and undirect structure learning algorithms for high dimensional complex data is, to our knowledge, new. First, we place the algorithms in the broader perspective of PNMs and analyse their learning method, the statistical criterion and presentationform (directed or undirected network). When it comes to evaluation of the algorithms, we will remove the confounding effect that the presentationform may have on network measures by transforming DAGs as learned by HC to UNMs. After, we evaluate the algorithms on topological and probabilistic accuracy paying attention to compactness and sparsity of the learned networks and veracious balance of different order dependencies.




\section*{Results}


\subsection*{BNs and PNs in the general framework of Probabilistic Gaussian Network Models}

\begin{figure}[ht]
\centering
		\includegraphics[width=\linewidth]{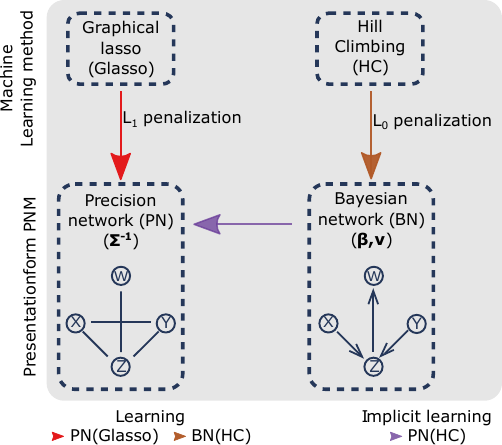}
		\caption{Schematic illustration of the presentationforms of Gaussian probability density function $\Prob$ in terms of conditional parameters $(\bm{\beta},\bm{\nu})$ and precision matrix $(\bm{\Sigma}^{-1})$, respectively associated with Bayesian Networks (BNs) and Precision Networks (PNs). Arrows refer to the associated learning algorithms (brown; score-based Hill Climbing algorithm, red; Graphical lasso together with the type of integrated statistical criterion ($l_0$ or $l_1$ penalization) and to the analytic transformation of the initially learned parameterset from BNs to PNs; PN(HC) (purple).}
\label{fig:scheme}
\end{figure}
\noindent Probabilistic Gaussian Network Models (PGNMs)\cite{Koller:2009:PGM:1795555} consist of graph and parameterset. The parameterset fully determines the associated Gaussian Joint Probability Density (JPD) function. 
The type of parameterset determines the presentationform of the graphical model, i.e. the `meaning' of the edges in the graph. A Gaussian JPD function can be determined by different types of parametersets such as the covariance matrix $(\bm{\Sigma})$, its inverse, the precision or concentration matrix $(\bm{\Sigma}^{-1})$, or linear regression coefficients in combination with local variation coefficients $(\bm{\beta},\bm{\nu})$.  Figure~\ref{fig:scheme} shows two learning methods and their associated presentationforms of the Gaussian JPD function: Precision Networks or pairwise Markov networks (PN $(\bm{\Sigma}^{-1})$) and Bayesian Networks (BN $(\bm{\beta},\bm{\nu})$).  In this study we leave out the analysis of Correlation Networks (CNs) that rely on the $(\bm{\Sigma})$ presentation  generally generated from a thresholded sample correlation matrix, because they cannot properly regulate high dimensional data. We refer the interested reader to Graafland et al. \cite{graafland_probabilistic_2020} for an extensive comparison study between BNs and CNs. The representations of the Gaussian JPD function in terms of $(\bm{\Sigma}^{-1})$) and $(\bm{\beta},\bm{\nu})$) are described in Methods section ``Probabilistic Gaussian Network Models" and subsections ``Probabilistic PN (BN) models".

As illustrated in Figure \ref{fig:scheme}, the algorithms under subject of study, Glasso (and modifications) and Hill climbing, differ in the parameterset (and hence the presentationform) they have as an objective to learn. 
They do however coincide on their machine learning spirit; both algorithms attempt to find a parameterset that optimizes a score function in which a statistical criterion (penalization) is integrated. The score functions and their statistical criteria are described in more detail in Methods sections ``learning PN (BN) structure from data". Learning methods (the steps in the algorithms) are chosen to efficiently execute this optimization task and are described in Methods section ``Learning with Hill Climbing and Glasso". 
In the case of Hill Climbing the search space is restricted to that of DAGs and in the case of Glasso to that of symmetric positive definite precision matrices, which, converted to binary format, are translated into undirected graphs. These constraints procure that the edges in the networks respectively are associated with the regression coefficients $\bm{\beta}$ in the $(\bm{\beta},\bm{\nu})$ parametrization of the JPD function and with the entries of the precision matrix $\bm{\Sigma}^{-1}$ in the $\bm{\Sigma}^{-1}$ parametrization. 

Once learned, a PNM is in general not bound to its initial parametrization/presentationform. At some cost, an analytical transformation of the parameterset can transform the presentationform of the PNM. In Figure~\ref{fig:scheme} the purple arrow represents a direct transformation between parametersets from BNs to PNs for which an analytical formula exists. The analytical transformation is described in Methods section ``Transformation of probabilistic BN model to probabilistic PN model". The transformation makes the initial BN loosing some information on its independency structure (see how the graphs of BNs and PNs encode independence statements about the JPD function in Methods section ``Dependencies in BN and PN structure"), but allows us to compare Hill Climbing and Graphical lasso in the same UNM framework.

\subsection*{Data, reference network and algorithm settings}\label{sec:ss.data}
We use two public available datasets for the Escherichia Coli genome to evaluate our structure learning algorithms with. On the one hand we use the (E.coli) micro-array dataset that is available from the Many Microbe Microarrays database (M3D) \cite{faith_many_2008} from which PNMs are learned with the four algorithms HC, Glasso, HGlasso and SFGlasso. In particular we use the latest version at this moment E\_coli\_v4\_Build\_6. This dataset contains the uniformly normalized expression levels of 4297 genes in E.coli measured under 466 experimental conditions using Affymetrix arrays. Expression levels from the same experimental conditions are averaged and their mean expression provides one of the 466 sample points in the dataset. 
A reference network, on the other hand, is constructed from a dataset containing evidence about transcriptional regulations; the interactions between Transcription Factors (TFs), that arise from TF encoding genes, and their target genes.
For the E.coli genome, a number of studies generated information on transcriptional regulations; The Regulon Data Base (RegulonDB) is the primary database \cite{santos-zavaleta_regulondb_2019} in which this information is gathered. We will use the most complete file in the RegulonDB called network\_tf\_gene.txt. 


We quire the algorithms to construct networks using only the expression levels in M3D of the 1683 genes from which evidence is reported in the RegulonDB. The known interactions in the RegulonDB will than really act like a reference network that us enables to evaluate the topological accuracy of the learned GRNs from the micro-array data. A similar strategy is applied in \cite{yu_enhanced_2017}. The 1683 genes contain a total of 173 TF encoding genes of which 172 count as the origin of all transcriptional regulation interactions in the reference network. Together they are good for 3381 unique interactions without self-loops. As interactions go from TF encoding genes to target genes the reference RegulonDB network is by nature a directed causal network.  
We expect the 172 TFs encoding genes in RegulonDB to (implicitly) determine the dependency structure of the M3D dataset. 

The number of edges that is needed to construct a PNM from the micro-array dataset plays an important role in the quality and the practical possibilities of a structure learning algorithm. With this in mind we generate networks of different sizes for all algorithms.
To vary the amount of edges $|\mathrm{E}|$ we vary for respectively Glasso, HGlasso and SFGlasso the initial parameters $\lambda$ and $\lambda_1, \lambda_2, \lambda_3$ and $\alpha$ (see Methods section ``Learning PN structure from data"). With respect to Hill Climbing we obtain networks of different amount of edges by varying the amount of iterations while using the standard $\BIC$ score (an action that give similar results as application of the $\BICg$ score and varying the parameter $\gamma$\cite{scutari_who_2019}.) 
The directed BNs learned by Hill Climbing will be transformed to UNMs. This process, described in Methods section ``Transformation of probabilistic BN model to probabilistic PN model", will cost some efficiency (extra, unnecessary parameters/edges will be added), but allow for comparison with the products of (SF-, H- and) Glasso as the transformed edges will encode the same type of parameters.

\subsection*{True Positives versus Network size}
\begin{figure}[ht]
\centering
		\includegraphics[width =\linewidth]{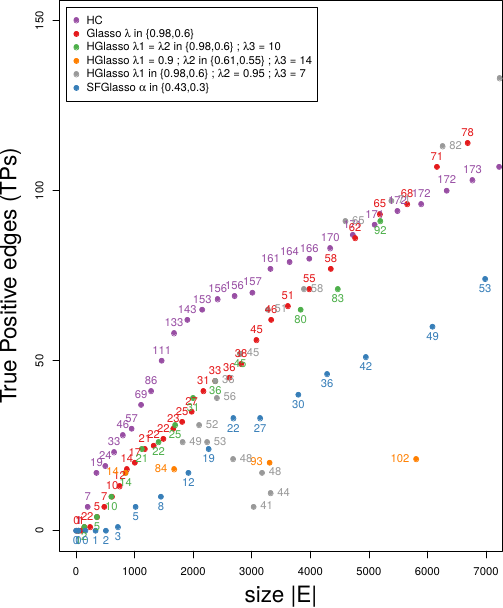}
		\caption{True Positives (dots) and amount of Transcription Factors (numbers) w.r.t. the networksize $|\mathrm{E}|$ as encountered by HC (purple), Glasso with $\lambda \in \{0.98,0.6\}$ (red), HGlasso with $\lambda_1 = \lambda_2 \in \{0.98,0.6\}, \lambda_3 = 10$ (green), HGlasso with $\lambda_1 = 0.9, \lambda_2 \in \{0.61,0.55\},  \lambda_3 = 14$ (orange),  HGlasso with $\lambda_1 \in \{0.98,0.6\}, \lambda_2 = 0.95, \lambda_3 = 7$ (grey), SFGlasso with $\alpha \in \{0.43, 0.3\}$ (blue).	}
		\label{fig:TP_TFs_vs_size}
\end{figure}

\noindent The topological accuracy of a single learned GRN is measured by the amount of True Positive edges (TPs) according to the reference RegulonDB network. We define an undirected edge in the learned UNMs as TP if there exists an associated directed edge in the RegulonDB network. 
Also, the amount of connected Transcription Factors (TFs) in the estimated network is measured as an indication of the hub structure of the learned networks. 
Mentioned above, these measures are analyzed in light of the networksize to measure the capability of algorithms to produce accurate, though compact and sparse networks. In Figure \ref{fig:TP_TFs_vs_size} results on TPs and TFs with respect to the number of total estimated edges $|\mathrm{E}|$ in the network are reported for the four algorithms. The dots indicate the amount of TPs and the dot labels, in the form of numbers, indicate the amount of TFs found in the belonging network.

The UNMs obtained with HC contain the highest amount of TPs and the highest amount of TFs for sparse networks up to 5000 edges. The biggest gain of TPs with respect to the amount of edges $|\mathrm{E}|$ occurs in the range from 1000 to 1800 edges. In this range also the amount of recovered TFs is highest (around 20 new TFs per 100 added edges). This simultaneous acceleration in grow is intuitive as all TPs in RegulonDB consist of at least one TF.
The classic Glasso recovers less TP edges as HC in sparse networks, the recovering rate is especially lower with respect to HC in the range of 1000 to 1800 edges, and could be related to the fact that in this range Glasso only recovers between 1 and 2 new TFs per 100 edges.  In bigger networks that consist of more than 5000 edges the amount of TPs exceeds HC values and we guess that in networks of around 20000 edges (not displayed) the amount of TFs will be almost equal. In the discussion we go into more detail into the strategy of Glasso and HC to discover TPs.

With respect to HGlasso, it depends on the network size and the combination of the values of the parameters $\lambda_1,\lambda_2$ and $\lambda_3$ if HGlasso can outperform Glasso or not.
Take for example the combination $\lambda_1 = 0.9$, $\lambda_2$ = variable and $\lambda_3 = 14$ (orange dots). These networks consist of more TFs as classic Glasso networks, but less TPs are found. The impact of the relaxed penalty term in the score function for hubs is thus visible, but not realized through the caption of more TPs by the HGlasso algorithm. The best results of HGlasso, are obtained with $\lambda_1$ = variable, $\lambda_2 = 0.95$ and $\lambda_3 =7$ (grey dots); medium networks (up from 3000 edges) with these parameters not only consist of more TFs, but also outperform the classic Glasso on TPs. 
Finally, we see that SFGlasso performs worse as the classic Glasso and HGlasso on both the amount of TPs and TFs. The discovery rate of TFs is slow resulting in little added TPs. It seems that the introduced scale-free degree distribution does not fit well to the dependency structure that is imposed by TFs in the micro-array data. 

\subsection*{Log-likelihood versus networksize }
\begin{figure}[ht]
\centering
		\includegraphics[width=\linewidth]{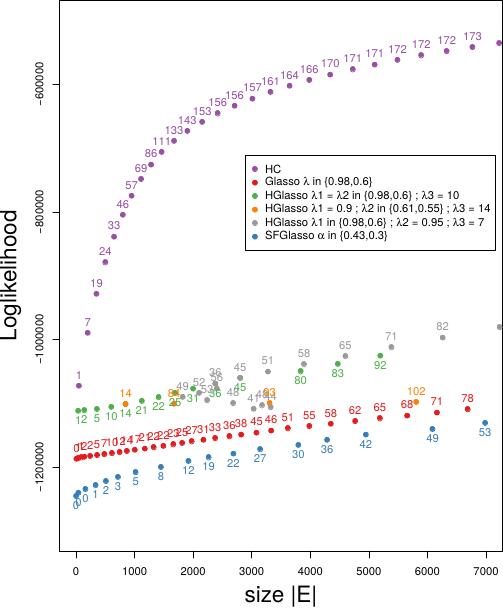}
		\caption{Loglikelihood values (dots) and amount of Transcription Factors (numbers) w.r.t. the networksize $|\mathrm{E}|$ as encountered by HC (purple), Glasso with $\lambda \in \{0.98,0.6\}$ (red), HGlasso with $\lambda_1 = \lambda_2 \in \{0.98,0.6\}, \lambda_3 = 10$ (green), HGlasso with $\lambda_1 = 0.9, \lambda_2 \in \{0.61,0.55\},  \lambda_3 = 14$ (orange),  HGlasso with $\lambda_1 \in \{0.98,0.6\}, \lambda_2 = 0.95, \lambda_3 = 7$ (grey), SFGlasso with $\alpha \in \{0.43, 0.3\}$ (blue).}
		\label{fig:logliks_TFs_vs_size}
\end{figure}

\noindent The relative probabilistic accuracy of a single learned GRN with respect to another learned GRN is measured by their difference in log-likelihood. The log-likelihood quantifies how well the model explains the data in terms of the likelihood of the available data given the particular probabilistic model (see Methods section ``Log-likelihood definition and calculation"). Figure \ref{fig:logliks_TFs_vs_size} displays the log-likelihood of the networks versus the amount of links in the networks. For easy comparison with Figure \ref{fig:TP_TFs_vs_size}, we again placed the number of integrated TFs as dotlabels. 
 
The log-likelihood curve of Hill Climbing rapidly improves until 2000 edges and establishes between 2000 and 7000 edges. Note that the same is true for the number of found TFs, growing rapidly until 2000 edges, reaching 153 TFs, and than slowing down until 7000 edges, reaching the total amount of 173 TFs. The true positives curve in Figure \ref{fig:TP_TFs_vs_size} also begins to flatten around 2000 edges. 
Hence, in the case of HC, the validation measures TP and TFs can be directly associated with the amount of information that is extracted from the micro-array data and captured in the network.

In the case of H-,SF- and Glasso the relation between log-likelihood and topological measures is more complex. The Glasso log-likelihood curve remains far apart from the HC curve for small and medium networks. From Figure \ref{fig:TP_TFs_vs_size} we learn that Glasso values of topological measures finally approach HC values when including more and more edges. 
Figure \ref{fig:logliks_TFs_vs_size} illustrates that this is not true (or in any case not true for regularized networks) for log-likelihood values. In the discussion we go into the relation between log-likelihood and TPs for Glasso and HC. 

All HGlasso networks score better on log-likelihood values than the classical Glasso. 
For HGlasso networks that improve Glasso networks on TPs (e.g. HGlasso networks with $\lambda_2 = 0.95$, $\lambda_3 = 10$ up from 3000 edges, grey dots) this betterment is even more pronounced. Still, however,  
HC log-likelihood values are not reached neither at small nor at medium edge size by the adapted HGlasso algorithm. 
Similar as in the case of TPs, SFGlasso scores worse than Glasso on log-likelihood values for small to medium networksizes. We do observe that the more edges are added to the SFGlasso network, the closer log-likelihood values become to Glasso values.

\subsection*{Illustration of networkstructure}
\begin{figure}[ht]
\centering
		\includegraphics[width=\linewidth]{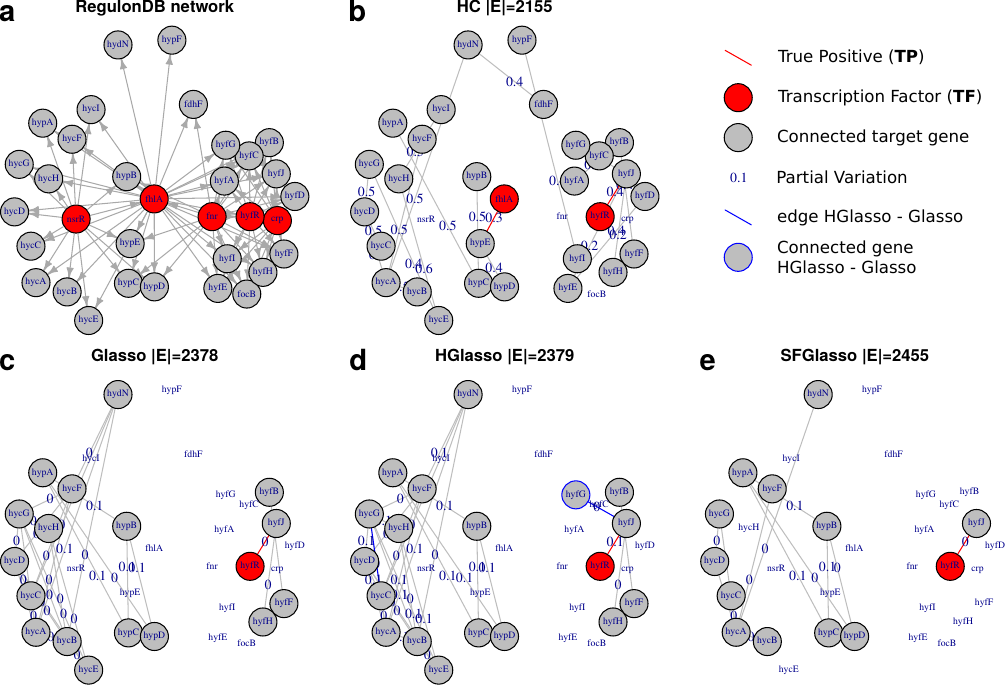}
		\caption{Reference and estimated networks: Window (\textbf{a}) is the part of the reference RegulonDB network that contains the two TF encoding genes hyfR and fhlA and their direct neighbours. Window (\textbf{b-e}) are the corresponding sub-networks of networks that are learned with respectively HC, (H)Glasso and SFGlasso. In window (\textbf{a}) TF encoding genes are coloured in red. In (\textbf{b-e}) TF encoding genes are coloured in red only when included in the network and connected genes are coloured grey. Red edges in (\textbf{b-e}) have there direct counterpart in window (\textbf{a}) (i.e. red edges are TPs).  In window (\textbf{d}) blue edges and blue vertex closure indicate edges and vertex that are included in the HGlasso network but not in the Glasso network of the same size, whereas grey edges and black vertex closures indicate edges and vertices that exist in both networks.}
		\label{fig:networks_TFs_hyfR_fhlA}
\end{figure}

\noindent To illustrate the above results on topological and probabilistic accuracy, we zoom into a part of the reference RegulonDB network that contains the two TF encoding genes hyfR and fhlA and we compare with the networks learned from M3D with the four structure learning algorithms. 
Window \textbf{a} in Figure \ref{fig:networks_TFs_hyfR_fhlA} shows the two TF encoding genes hyfR and fhlA  (coloured in red) and their direct neighbours in the RegulonDB network (among the direct neighbours are three other TFs that are also colored in red; crp, fnr and nsrR). The RegulonDB is a directed network in which the meaning of the edges is causal.
Symmetric dependencies that could exist between childs of hyfR and/or fhlA due to their common parent are not documented in RegulonDB and hence are not displayed in window \textbf{a}. 
Windows \textbf{b}-\textbf{e} show the same subset of nodes as window \textbf{a} and form part of networks that are learned from M3D by respectively HC, Glasso, HGlasso and SFGlasso. These networks are all (converted to, in the case of HC,) undirected PNMs (or UNMs) in which edges do not indicate casual influence, but direct dependency (the edges represent non zero entries in the -symmetric!- precision matrix). The exact encoding of dependencies in the networks is described in the Methods section ``Dependencies in PN". The grey nodes indicate genes that are integrated in the network and red edges indicate edges that find their direct counterpart in the reference network in window \textbf{a}, i.e. edges that are denominated TPs. The edge-labels display the weight of the edges, the estimated partial variation. In window \textbf{d} blue edges and blue vertex closure indicate edges and vertex that are included in the HGlasso network but not in the Glasso network in window \textbf{c} of similar size, whereas grey edges and black vertex closures indicate edges and vertices that exist in both networks. 

The results in the above subsections are illustrated in the subnetworks. 
In a sparse network, Hill Climbing (window \textbf{b}) discovers both TF encoding genes, hyfR and fhlA and with them two TP edges between respectively hyfR and hyfJ and fhlA and hypE. Glasso, HGlasso and SFGlasso (window \textbf{c},\textbf{d} and \textbf{e}) discover only one TF, hyfR, and find one TP related to hyfR.

Despite finding the most TPs, Hill climbing also discovers the most unique child-child dependencies between the target genes that are regulated by the transcription factors HyfR and FhlA. Child-child dependencies are not considered as TPs in the RegulonDB network, but do improve quality of UNM networkstructure. For example, the community $\{\mathrm{hyfA, hyfB} \dots \mathrm{hyfJ}\}$ and the TP between hyfR and hyfJ can be an indication that the set $\{\mathrm{hyfA}, \mathrm{hyfB} \dots \mathrm{hyfJ}\}\setminus\{\mathrm{hyfJ}\}$ is also progeny of HyfR. 

Glasso finds less variety of child-child dependencies between nodes that are regulated by HyfR and FhlA. Instead, the sub groups of childs are interconnected with more edges between them. These edges are probably true, but, from an information theoretic perspective, are redundant in the sense that they lower entropy of the community structure in the network (see also \cite{graafland_probabilistic_2020}). Moreover, the values of the parameters, the partial variation, in Glasso differ in magnitude with the values of the parameters in HC. The low partial variation as estimated by (H)Glasso lower loglikelihood values with respect to Hill Climbing. In the discussion we will come back to the estimation of the magnitude of the parameters and its impact on log-likelihood values. 

In window \textbf{d}, HGlasso improves the networkstructure with respect to Glasso with an additional connected gene hyfG while using the same amount of edges. Also, parameter estimation seems more accurate as HGlasso includes higher partial variation values.  In general, the inclusion of more significant edges (for some combinations of $\lambda_1,\lambda_2,\lambda_3$) and more precise parameter estimation alter the log-likelihood values of HGlasso with respect to Glasso, showed in Figures \ref{fig:TP_TFs_vs_size} and  \ref{fig:logliks_TFs_vs_size}.

The SFGlasso in window \textbf{e} on the other hand excludes four more genes than the classic Glasso. Not many dependencies are found in this part of the sub network. Thereby, the estimation of parameters is worse in this part of the sub network. Partial variations are of even smaller magnitude with respect to Glasso. These results are in accordance with Figure \ref{fig:TP_TFs_vs_size} showing that SFGlasso networks are less complete than (H)Glasso networks and with Figure \ref{fig:logliks_TFs_vs_size} showing that the least amount of information can be induced from SFGlasso networks. 

\section*{Discussion}

Our results on sparse and medium networks showed high log-likelihood values for HC with respect to the classic Glasso and its modifications. Morever, the results for sparse networks (up to 5000 edges) show that the HC algorithm finds the most true interactions (in the form of TPs). We first discuss the log-likelihood results in light of the topology of the learned networks and in light of the score functions that are integrated in the algorithms. Later we discuss the results on TPs in light of the same score functions. In the second part of the discussion we place the results of HC in the light of two other partial-variation learning algorithms, SPACE and ESPACE, that are evaluated in \cite{yu_enhanced_2017} on the same E.Coli dataset. In the third part we discuss the results of HC in this study with the results of HC in a recent study on high-dimensional climate data and find some generalities about Gaussian BNs that are learned with HC and applied on high-dimensional real world data\cite{graafland_probabilistic_2020}. 

When analyzing log-likelihood values, firstly we should keep in mind that not all interactions that can be inferred from the M3D are equally informative and thus not contribute equally to the log-likelihood. It is intuitive to regard the evidence in the RegulonDB that takes the form of TF-target gene interactions as the backbone structure of the M3D data, whereas local interaction structure between childs of the same TPs is less informative with respect to the complete M3D database. 
In this light, the HC algorithm produces network topologies that contribute to high loglikelihood values, finding relatively more TPs and much more TFs than Glasso and its modifications in sparse networks and just enough links to determine local interaction structure. 
Secondly, accurate weight estimation of the incorporated paramaters (or the weight of the edges) alters loglikelhood values. The HC algorithm is able to accurately estimate the weight of parameters. This is fruit of the $l_0$ penalty used in the score function that only penalizes the \textit{amount} of parameters and not their weight. The $l_1$ regularization in Glasso, on the other hand, while penalizing the amount of parameters, also penalizes the \textit{weight} of parameters. The $l_1$ penalty thus imposes a structural bias on the selected parameters in the JPF, resulting in inaccurate parameter estimation and substantially lower log-likelihood values. The score functions including $l_0$ or $l_1$ penalization are given in detail in Methods section ``Learning BN (PN) structure from data". 

The answer to the question why HC includes more TPs and TFs than Glasso and modifications has to do with the integrated score functions and with the steps in the learning algorithms. Every iteration HC can choose to incorporate a parameter. This edge is selected in order to maximize the score function that partly exists of the log-likelihood and partly of the $l_0$ penalization term that penalizes the amount of parameters. On the one hand there are high informative links (TPs) from which there are relative few that alter the log-likelihood substantively. On the other hand, there are local links, from which there are many that are relatively less informative. HC is designed to alter log-likelihood with the least amount of parameters and thus includes first TPs and informative local links (that enlarge communities) and only then fills the network with less informative local links (that fill communities). 
The score function of Glasso on the other hand exists of the same log-likelihood part and of the $l_1$ penalization term that penalizes besides the amount also the weight of edges. Local links have a stronger direct correlation and it is for this reason that Glasso includes more of them, whether informative or uninformative, as local links `remain' under application of the statistical criterion. More informative links as TPs have weaker direct correlation and in sparse networks Glasso shrinks those links to zero.  

In the same line the differences in results between Glasso and modifications can be explained. Due to the adapted $l_1$ regularization, HGlasso performs better on log-likelihood values than the classical Glasso for every combination of initial parameters $\lambda_1,\lambda_2, \lambda_3$. The new score function always alleviates the penalization on parameters that are related to hubs, resulting in higher log-likelihood values with respect to the classic Glasso. HGlasso, however, still depends heavily on $\lambda_1$ (the penalty term for non hubs) to introduce overall sparsity, and thus can not reduce the mean parameter bias sufficiently to reach HC values for small and medium networks.
With respect to the measure of true positives, it depends on the combination of tuning parameters and on the networksize if HGlasso performes out Glasso. A successful combination of tuning parameters leads HGlasso finding more TPs. These networks generally contained more TFs than the associated Glasso network. 
However, the incorporation of more TFs did not always led to more True Positives in the network. We found no standard method to discover combinations of initial parameters with whom Glasso could be outperformed. Finding such a combination is a time consuming process of trial and error. 

The algorithm SFGlasso does incorporate a hub structure, but the networks generally contain less TFs than Glasso and the combinations of HGlasso. As judged by the low log-likelihood values, the algorithm thus selects `wrong' hubs and also places great bias on the estimates of informative links. These results may have to do with the poor adjustment of SFGlasso to high dimensional data. Generally, in scale free networks, the coefficient $\alpha$ in the powerlaw with which the degree distribution decays lies between 2 and 3. However, we had to reduce $\alpha$ to values below 0,5 as otherwise the resulting networks stayed empty (even when using only two iterations). Moreover, in this work we only display results after two iterations, because using more than two iterations the lasso algorithm took extraordinary long time to converge. 

The analysis of HGlasso and SFGlasso with respect to Glasso is a strong indication that GRNs are not fully characterized by scale-free degree distributions and/or hub nodes. The structure of a GRN seems to be more complex and probably not completely definable. We expect algorithms that focus only on these specific characteristics while leaving out other characteristics of the real-world data to improve the classic Glasso only up to certain height. \\

Mentioned before two other algorithms were applied on (a less up to date version of) the micro-array data used in this paper: the algorithms SPACE and ESPACE\cite{yu_enhanced_2017}. The algorithm SPACE uses sparse covariance selection techniques but differs from Glasso and was especially designed for the large-$p$-small-$n$ data in the framework of GRNs and for powerful identification of hubs\cite{peng_partial_2009}. The ESPACE algorithm is an extension of the SPACE method and explicitly accounts for prior information on hub genes, which, in case of E.Coli data, yields knowing in advance that TFs are the highly connected nodes in the true E.Coli GRN. 
We can loosely compare the HC algorithm with the performance of the algorithms SPACE and ESPACE. 
In table 6 of their work, Yu et al.\cite{yu_enhanced_2017} report a percentage of 4.35\% of TPs in a network of 386 edges for SPACE, whereas ESPACE found 12,89\% TPs in a network of 349 edges. In our work HC has a percentage of 4.85\% of TPs in a network of 350 edges, outperforming SPACE. We may conclude with some precaution, as datasets are not entirely equal, that HC performs out the exploratory algorithm SPACE, but can not compete with ESPACE that starts with prior information about the true hubs.\\



Finally, the great power of HC when exploring E.coli data with unknown structure is that one can extract information of the true deterministic underlying network structure from every edge, as was illustrated at hand of the graphs in Figure \ref{fig:networks_TFs_hyfR_fhlA}. The fact that no edge is redundant with respect to the dataset enables us to learn from every `false' positive edge, whereas in the case of Glasso and variants this is not true: not every edge is data-significant neither tells us more about the network structure. 
The same conclusion was drawn for high dimensional climate data for which we found that HC provided an informative community structure that can be analyzed well with centrality measures (i.e. betweenness centrality, see \cite{graafland_probabilistic_2020})). The HC approach prioritizes an efficient edge distribution by favouring heterogeneous selection of edges between communities over uniform selection of edges that lie in communities. We saw that this approach explores very well deterministic features in high dimensional real world complex systems like interconnected spatial communities by teleconnections (climate) or regulatory interactions (in GRNs). Complex network centrality measures such as community structure and betweenness centrality therefore have high potential for probabilistic BN networks applied on real-world datasets.

\section*{Conclusions}
The use of the Hill Climbing algorithm that arises in the context of Gaussian Bayesian networks offers a sound approach for the reconstruction of gene regulatory networks of high-dimensional-low-samplesize micro-array data when no initial information is at hand about the underlying complex dependency structure.
The HC algorithm picks only the most significant dependencies from the Gaussian data and in this manner naturally includes the complex dependency structure of the complex GRN that may consist of hubs, a scale-free degree distribution, outliers, a combination of the former or of other real-world characteristics. The algorithm naturally leaves out uninformative dependencies and variables.

The Bayesian network can easily be transformed to an Undirected Gaussian Probabilistic Network Model, paying the price of some loss of information on the independence structure with respect to its initial directed Bayesian Network. 
If one prefers an undirected PNM over a directed PNM -- for easy interpretation due to symmetric links, or for sparse estimation of the inverse covariance matrix -- this study shows that the transformation from BN to UNM is worth this loss of information as the UNMs obtained by Hill Climbing still outperform UNMs obtained by state of the art UNM-structure learning algorithms when applied to high-dimensional-low-sample size  (micro-array) data that contains an unknown complex dependency structure.

This conclusion is drawn with respect to state-of-the-art structure algorithms that arise in the context of Undirected Gaussian Network Models, the Glasso algorithm and variants of Glasso that are developed to integrate complex dependency structure. These algorithms model unnecessary dependencies at the expense of the probabilistic information in the network and of a structural bias in the probability function that can only be relieved including many parameters. In the case of the E.Coli gene expression data used in this work, unnecesary dependencies also go at the expense of the amount of true positive edges, the last as judged by a reference network compounded of evidence gathered in the RegulonDB.

\section*{Methods}

\subsection*{Probabilistic Gaussian Network Models (PGNMs)}
The term refers to the choice of a multivariate 
Gaussian Joint Probability Density (JPD) function to associate graph
edges with model parameters in a given PNM, such that
the probabilistic model encodes in the JPD function a large number of 
random variables that interact in a complex way with each other by a 
graphical model. 
The multivariate Gaussian 
JPD function can take various representations in which dependencies between the variables 
are described by different types of parameters. The best-known representation of the Gaussian JPD function is in terms of 
marginal dependencies, {\it i.e.}, dependencies of the form $X_i,X_j|\emptyset$
as present in the covariance matrix $\bm{\Sigma}$. Let $\X$ be a $N$-dimensional 
multivariate Gaussian variable then its probability density function 
$\Prob(\X)$ is given by:
\begin{equation}
	\label{eq:mvg-cov}
	\Prob(\X) = (2\pi)^{-N/2}\det(\bm{\Sigma})^{-1/2}\exp \{-1/2(\X-\bm{\mu})^\top\bm{\Sigma}^{-1}(\X-\bm{\mu})\},
\end{equation}
where $\bm{\mu}$ is the $N$-dimensional mean vector and $\bm{\Sigma}$ the 
$N \times N$ covariance matrix. In the following we describe in some detail two types of PGNMs, in which 
parameters reflect respectively direct
dependencies $X_i,X_j|\X\backslash\{X_i,X_j\}$ and general conditional dependencies $X_i|\mathcal{S}$ with 
${\mathcal S}\subseteq \mathbf{X}$ (direct 
dependencies are the least restrictive case of conditional dependencies).

\subsection*{Probabilistic PN models}
The Gaussian JPD function in equation (\ref{eq:mvg-cov}) can be formulated more generally using a set of factors $\Phi = \{\phi_1(\mathcal{S}_1),\dots,\phi_k(\mathcal{S}_k)\}$ that describe dependencies between arbitrary (overlapping) subsets of variables $\mathcal{S}_1,\dots,\mathcal{S}_k$ which comply with $\cup_k \mathcal{S}_k = \X$.
This representation of the JPD is called the Gibbs function and written as \cite{Koller:2009:PGM:1795555}
\begin{equation}
\label{eq:gibbs1} 
\Prob(X_1,\dots,X_N) = \frac{1}{Z}\tilde{\Prob}(X_1,\dots,X_N),
\end{equation}
with
\begin{equation}
\label{eq:gibbs2} 
\tilde{\Prob}(X_1,\dots,X_N) = \prod_{i=1}^{k} \phi_i(\mathcal{S}_i) \hspace{10pt} \text{and} \hspace{10pt} Z = \sum_{i=1}^{N}\tilde{\Prob}(X_1,\dots,X_N).
\end{equation}

The Gibbs distribution where all of the factors are over subsets of single variables or pairs of variables is called a pairwise Markov network. The factors in a pairwise Markov network correspond to direct dependencies, {\it i.e.}, dependencies of the form $X_i,X_j|\X\backslash\{X_i,X_j\}$. In a Gaussian distribution these dependencies are present in the inverse covariance matrix or precision matrix $\bm{\Sigma}^{-1}$.
The information form of the Gaussian JPD function in terms of the precision matrix $\bm{\Theta} = \bm{\Sigma}^{-1}$ 
\begin{equation}\label{eq:mvg-prec}
\Prob(\X, \mu = 0, \bm{\Theta}) = (2\pi)^{-N/2}|\bm{\Theta}|^{1/2}\exp \{-1/2\sum_{i}\theta_{ii}X_i^2-\sum_{i<j}\theta_{ij}X_iX_j\}.
\end{equation}
is equivalent to the Gibbs function in equation (\ref{eq:gibbs1}) with factors defined on every variable and every pair of variables, i.e. $\Phi = \{\Phi^n,\Phi^e\}$ with  $\phi^n_i = \exp\{-\frac{1}{2} \theta_{ii}X_i^2\}$ and $\phi^e_{ij} = \exp\{\theta_{ij}X_i X_j\}$.

The corresponding PGNM in which the notion of variables and pairs of variables is extended to the notion of nodes and undirected edges in a graph is called the Probabilistic PN model. 
The graph of a PN encodes the 
probability function in equation (\ref{eq:mvg-prec}) as follows. Each node 
corresponds to a variable $X_i \in \X$, the presence of an edge 
$X_i - X_j$ implies the presence of the factor 
$\phi^e_{ij}$ in $\Prob(\X)$, and direct 
dependency of $X_i$ and $X_j$. Moreover, the absence of an arc between 
$X_i$ and $X_j$ in the graph implies the absence of the factor $\phi^e_{ij}$
in $\Prob(\X)$ and, thus, the existence of a set of variables   
${\mathcal S} \subseteq \X\backslash\{X_i,X_j\}$ that makes $X_i$ and $X_j$  
conditionally independent in 
probability~\cite{Koller:2009:PGM:1795555}.

The graph structure of PNs in this work are estimated simultaneously with the values of the parameters in $\bm{\Theta}$ that define this structure. This simultaneous learning process is explained in Methods section ``Learning PN structure from data". 

\subsection*{Probabilistic BN models}
Alternatively, the $\Prob(\X)$ in equation~(\ref{eq:mvg-cov}) can be 
characterized with conditional dependencies of the form $X_i|\mathcal{S}$ with 
${\mathcal S}\subseteq \mathbf{X}$. The representation of the JPD is then a product of 
Conditional Probability Densities (CPDs):
\begin{equation}
	\label{eq:mvg-fac} 
	\Prob(X_1,\dots,X_N) = \prod_{i=1}^{N} \Prob_i(\XPi)
\end{equation}
with
\begin{equation}
	\label{eq:mvg-loc} 
	\Prob(\XPi) \sim {\cal N}\left(\mu_i + 
	\sum_{j|X_j \in \Pi_{X_i}}\beta_{ij}(X_j-\mu_j), \; \nu_i\right)
\end{equation}
whenever the set of random variables $\{\XPi\}_{i\in N}$ is 
independent\cite{shachter_gaussian_1989}. In this representation ${\cal N}$ 
is the normal distribution,
$\mu_i$ is the unconditional mean of $X_i$, $\nu_i$ is the conditional 
variance of $X_i$ given the set $\Pi_{X_i}$ and $\beta_{ij}$ is the 
regression coefficient of $X_j$, when $X_i$ is regressed on $\Pi_{X_i}$. 
We call $\Pi_{X_i}$ the parentset of variable $X_i$.

The corresponding PGNM in this case is the Probabilistic BN model. 
The graph of a BN model is a $\DAG$ encoding the 
corresponding probability distribution as in equation~(\ref{eq:mvg-fac}). Each node 
corresponds to a variable $X_i \in \X$, the presence of an arc 
$X_j \rightarrow X_i$ implies the presence of the factor 
$\Prob_i(X_i|\dots X_j \dots )$ in $\Prob(\X)$, and thus conditional 
dependence of $X_i$ and $X_j$. Moreover, the absence of an arc between 
$X_i$ and $X_j$ in the graph implies the absence of the factors 
$\Prob_i(X_i|\dots X_j \dots )$ or $\Prob_j(X_j|\dots X_i \dots )$ 
in $\Prob(\X)$ and, thus, the existence of a set of variables   
${\mathcal S} \subseteq \X\backslash\{X_i,X_j\}$ that makes $X_i$ and $X_j$  
conditionally independent in 
probability~\cite{Koller:2009:PGM:1795555,castillo_expert_1997}.

The graph structure of the BN identifies the parentset $\Pi_{X_i}$ 
in equation~(\ref{eq:mvg-fac}). With this structure available, one easily learns 
the corresponding parameter set $(\bm{\beta},\bm{\nu})$; in our case parameters 
$\beta_{ij}$ and $\nu_i$ are a maximum likelihood fit of the linear 
regression of $X_i$ on its parentset $\Pi_{X_i}$. To estimate parameter values from graph structure we use the appropriate 
function in the R-package {\sc bnlearn}  \cite{scutari_learning_2010}.
The challenge of learning the graph structure is explained in Methods 
section ``Learning BN structure from data".

\subsection*{Learning PN structure from data}
A Precision Network (PN) is learned with the help of a structure learning 
algorithm that estimates the inverse covariance matrix, i.e. the precision matrix $\bm{\Sigma}^{-1}$ of the underlying Gaussian distribution. Converted into binary format, the estimate $\bm{\Theta}$ of $\bm{\Sigma}^{-1}$ provides the undirected adjacency matrix $\bm{A}$ of a pairwise Markov Network. 
From the adjacency matrix $\bm{A}$ of the graph of a PN the structure of the factor-set $\Phi = \{\Phi^n,\Phi^e\}$ of the associated Gaussian JPD function (outlined in Eq.~(\ref{eq:mvg-prec})) can be directly read off. In the Methods Section ``Probabilistic PN Models" is explained how pairwise Markov networks encode the corresponding Gaussian PNM.  \\

The Graphical lasso (Glasso) can be regarded as the default structure learning algorithm learning PNs for large-$p$-small-$n$ datasets. Glasso is a score-based algorithm based on a convex score. This score is basically made up by the Maximum Likelihood Estimate (MLE) of the precision matrix of a Gaussian probability function to which an $l_1$ penalty term is added\cite{friedman_sparse_2008}:
\begin{equation}\label{eq:score_lasso}
    \Score(\bm{\Theta}, \S,\lambda) = \log(\det{\S\bm{\Theta}}) - tr(\S\bm{\Theta}) - \lambda \|\bm{\Theta}\|_1.
\end{equation}
Here $\bm{\Theta} = \bm{\Sigma}^{-1}$, $\S$ is the sample covariance matrix calculated directly from the data $\D$ and $\lambda$ a scalar, the penalization coefficient. 
Networks of different sizes can be generated by varying the penalization parameter $\lambda$. In this work we generate networks with zero edges to complete networks by varying $\lambda$ from 1 to 0. A short outline of the steps in the Graphical lasso algorithm is given in Methods section ``Learning with Hill Climbing and Glasso". The Glasso function is implemented in the R-package {\sc glasso} \cite{friedman_sparse_2008}. \\

The Hub Graphical lasso (HGlasso) learns a PN that consist of hub nodes combining a lasso ($l_1$) penalty and a sparse group lasso ($l_2$) penalty \cite{tan_learning_2014}. The estimated inverse covariance matrix $\bm{\Theta}$ can be decomposed as $\bm{\Theta} = \mathbf{Z} + \mathbf{V} + t(\mathbf{V})$, where $\mathbf{Z}$ is a sparse matrix and $\mathbf{V}$ is a matrix that contains hub nodes. The belonging score is
\begin{eqnarray}
\Score(\bm{\Theta}, \S,\lambda_1,\lambda_2,\lambda_3) & = & \log(\det{\S\bm{\Theta}}) - tr(\S\bm{\Theta}) \nonumber\\
& & - \lambda_1 \|\mathbf{Z}\|_1 - \lambda_2\|\mathbf{V} - \text{diag}(\mathbf{V})\|_1 \nonumber\\
& & -\lambda_3\sum_{j = 1}^p \|(\mathbf{V} - \text{diag}(\mathbf{V}))_j\|_2,
\end{eqnarray}
with $\bm{\Theta}$ restricted to $\bm{\Theta} =\mathbf{V}+\mathbf{V}^T+\mathbf{Z}$. In this score $\lambda_3$ controls the selection of hub nodes, and $\lambda_2$ controls the sparsity of each hub node’s connections to other nodes. We obtain networks of different sizes by varying $\lambda_1,\lambda_2$ and $\lambda_3$. The HGlasso function is implemented in the R-package {\sc hglasso} \cite{tan_learning_2014}. \\

The Scale-Free Graphical lasso (SFGlasso) aims to include even more structural information than mere sparsity or hubs \cite{liu_learning_2011}. Hubs are expected in this type of network but the focus lies on learning models that poses the so-called “scale-free” property; a property often claimed to appear in real-world networks. This feature is mathematically expressed by a degree distribution $p(d)$ that follows a powerlaw: $p(d)\propto d^{-\alpha}$ (up from a certain degree $d$). In the score function of the classic Glasso, the $l_1$-edge regularization is replaced with a power law regularization. The objective score function is:
\begin{eqnarray}
\Score(\bm{\Theta}, \S,\alpha,\beta) & = & \log(\det{\S\bm{\Theta}}) - tr(\S\bm{\Theta}) \nonumber\\ 
& & -\alpha\sum_i\log(\|\bm{\Theta}\neg i\|_1 +\epsilon_i)
- \beta\sum_i|\theta_{ii}|,
\end{eqnarray}
with $\bm{\Theta}_{\neg i} = \{\theta_{ij}|j\neq i\}$.
This score function is not convex, a requirement to use Glasso, however can be proven to be monotone increasing. The score $\Score(\bm{\Theta}, \S,\alpha,\beta)$ is sequentially improved by elements of the sequence $\bm{\Theta}^{n}$ that iteratively maximize the following reweighted convex $l_1$ regularization problems:
\begin{eqnarray}\label{eq:sfglasso2}
\Score(\bm{\Theta}^{n+1},\S,\lambda_{ij}) & = & \log(\det{\S\bm{\Theta}^{n+1}}) - tr(\S\bm{\Theta}^{n+1})\nonumber \\
& & - \sum_{i\neq j}\lambda_{ij}|\theta^n_{ij}|-\beta\sum_i|\theta^n_{ii}|,
\end{eqnarray}
where $\lambda_{ij} = \alpha(\frac{1}{\|\bm{\Theta}^n_{\neg i}\|_1+\epsilon_i}+\frac{1}{\|\bm{\Theta}^n_{\neg j}\|_1 + \epsilon_j}).$
This re-weighting reduces regularization coefficients of nodes with high degree, encouraging the appearance of hubs with high degree. 

Following the set up in the experiment section of \cite{liu_learning_2011} we take $\beta_i = 2\alpha/\epsilon_i$ and $\epsilon_i$ equal to $\theta_{ii}$ estimated  in  the  last iteration,  in this way  $\epsilon_i$ is  on  the  same  magnitude of $\|\bm{\Theta}^n_{\neg j}\|_1$. Generally, in scale free networks $\alpha$  lies between 2 and 3. However, we had to reduce $\alpha$ to values below 0,5 as otherwise the resulting networks stay empty (even when only using 2 iterations). To optimize equation (\ref{eq:sfglasso2}) and find $\bm{\Theta}^{n+1}$ we iteratively use the glasso function in the R-package {\sc glasso} with $\lambda = \lambda(\bm{\Theta}^{n})$, defined above.
In this work we display results after 2 iterations. Using 3-5 iterations the lasso algorithm took extraordinary long time to converge.  

\subsection*{Learning BN structure from data}
The graph of a BN is estimated  with the help of a structure learning algorithm that finds the conditional dependencies between the 
variables and encodes this information in a DAG. 
Graphical (dis-)connection in the DAG implies conditional 
(in-)dependence in probability (see Methods section ``Dependencies in BN"). From the structure of a BN a factorization 
of the underlying JPD function $\Prob(\X)$ of the multivariate random 
variable $\X$ (as given by Eq.~(\ref{eq:mvg-fac})) can be deduced. In the Methods Section ``Probabilistic 
BN Models" is explained how networks can be extended 
to their corresponding Probabilistic Network Model (PNMs). 

In general there are three types of structure learning algorithms: 
constrained-based, score-based, and hybrid structure learning algorithms--- 
the latter being a combination of the first two algorithms.

Constrained-based algorithms use conditional independence tests of the 
form $\mathrm{Test}(X_i,X_j|\mathcal{S};\D)$ with increasingly large candidate separating 
sets $\mathcal{S}_{X_i,X_j}$ to decide whether two variables $X_i$ and $X_j$ are 
conditionally independent. All constraint-based algorithms are based 
on the work of Pearl on causal graphical 
models~\cite{Verma:1990:ESC:647233.719736} and its first practical 
implementation was found in the Principal Components 
algorithm~\cite{spirtes_causation_1993}. 
In contrast, score-based algorithms apply general machine learning 
optimization techniques to learn the structure of a BN. Each candidate 
network is assigned a network score reflecting its goodness of fit, 
which the algorithm then attempts to maximise~\cite{russell_artificial_1995}.
In~\cite{scutari_who_2019} we
compared structure learning 
algorithms belonging to the three different classes on accuracy and 
speed for high-dimensional complex data. 
We found that score-based algorithms perform best. Algorithms in this class are able 
to handle high-variable-low-sample size data and find networks of all 
desired sizes. Constrained-based algorithms can only model complex data 
up to a certain size and, as a consequence, for climate data they only reveal
local network topology. Hybrid algorithms perform better than 
constrained-based algorithms on complex data, but
worse than score-based algorithms.

In this work we use a simple score-based algorithm, the Hill Climbing 
(HC) algorithm~\cite{russell_artificial_1995}, to learn BN structure.
The HC algorithm starts with an empty graph and iteratively adds, removes 
or reverses an edge maximizing the score function. This algorithm is formalized in Methods section “Learning with Hill Climbing and Glasso”. HC is implemented in the R-package {\sc bnlearn}.

We used the Bayesian 
Information Criteria (BIC) (corresponding to $\BIC_0$ in 
\cite{scutari_who_2019}) score, which is defined as: 
\begin{equation}
	\BIC(\G; \D) =
	\sum_{i=1}^N \left[\; \log \Prob(\XPi) - \frac{|\T|}{2}\log N \;\right],
	\label{eq:bic}
\end{equation} 
where $\G$ refers to the graph (DAG) for which the BIC score is calculated, P refers to the probability density function that can be deduced from 
the graph (see Methods Section Probabilistic BN 
Models.), $\Pi_{X_{i}}$ refer to the parents of $X_i$ in the graph 
({\it i.e.} nodes Y with relation $Y \rightarrow X_i$ in the graph) and 
$|\bm{\Theta}_{X_{i}}|$ is the amount of parameters of the local density function $\Prob(\XPi)$. 

\subsection*{Dependencies in BN and PN structure}
\begin{figure}[ht]
\centering
		\includegraphics[width=\linewidth]{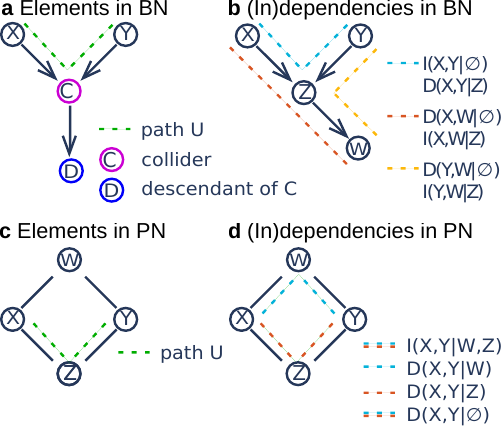}
		\caption{(\textbf{a}) and (\textbf{c}): Nomenclature of elements in respectively a Bayesian Network (BN) and a Precision Network (PN). (\textbf{b}) and (\textbf{d}): Some (in)dependencies in simple BN and PN consisting 
			of four nodes $X$, $Y$, $Z$ and $W$. In (\textbf{b}) two sets of nodes are dependent given a third if conditions (1) and (2) in the main text are fulfilled. 
			On the one hand, the conditional 
			relationship $X,Y|Z$ and the marginal relationships 
			$X,W|\emptyset$ and $Y,W|\emptyset$ satisfy conditions (1) and (2), 
			so that we have $\mathrm{D}(X,Y|Z)$, $\mathrm{D}(X,W|\emptyset)$ and $\mathrm{D}(Y,W|\emptyset)$. 
			On the other hand, the marginal relationship $X,Y|\emptyset$ violates 
			condition (1) and the conditional relationships $X,W|Z$ and $Y,W|Z$ 
			violate condition (2), so that we have $\mathrm{I}(X,Y|\emptyset)$ and $\mathrm{I}(X,W|Z)$ and $\mathrm{I}(Y,W|Z)$.  In (\textbf{d}) the conditional relationships $X,Y|W$, $X,Y|Z$ and $X,Y|\emptyset$ satisfy the condition of graphical dependence in a PN, and hence the statements $\mathrm{D}(X,Y|W)$, $\mathrm{D}(X,Y|Z)$ and $\mathrm{D}(X,Y|\emptyset)$ hold. On the other hand the conditional relation $X,Y|W,Z$ does not satisfy the condition of graphical dependence; there does not exist a path $\mathrm{U}$, such that neither $W$ nor $Z$ is not on $U$. Thus $\mathrm{I}(X,Y|W,Z)$ holds.
			\newline\textbf{Formal proof of} $\mathbf{D}(\bm{X,Y|Z})$ \textbf{in (b)- Conditional dependence of $X$ and $Y$ given $Z$.} 
			The conditioning set $\mathcal{S}$ exists of $Z$. The only path between $X$ and $Y$ is the blue path. Hence we declare the blue path U. $Z$ is a collider and $Z$ is in $\mathcal{S}$. 
			There are no other colliders on U. Hence condition (1) is satisfied. $Z$ is the only variable on U. And $Z$ is a collider. 
			Thus, U does not contain non-colliders. Hence condition (2) is satisfied. 
			As condition (1) and (2) are satisfied we have that $X$ and $Y$ are dependent given $Z$, i.e.~$\mathrm{D}(X,Y|Z)$.}
			\label{fig:dependencies}
\end{figure}

\noindent In the following we describe how a BN (or DAG) and a PN (or Pairwise Markov Network) encode
conditional dependencies. New nomenclature is indicated with an asterisk and illustrated in Figure~\ref{fig:dependencies}a and c.
\subsection*{Dependencies in BN}
In a BN two nodes $X$ and $Y$ are conditionally dependent given a set $\mathcal{S}$ (denoted by $\mathrm{D}(X,Y|\mathcal{S})$) if and only 
if they are graphically connected, that is,
if and only if there exists a path $\text{U}^*$ between $X$ and $Y$ satisfying the following two
conditions: 
\begin{itemize}
	\item {\em Condition (1)}: for 
	every collider$^*$ C (node C such that the part of U that goes over C has the 
	form of a V-structure, {\it i.e.},  $\rightarrow C  \leftarrow$) on U, either 
	C or a descendant$^*$ of C is in $\mathcal{S}$.
	\item {\em Condition (2)}: no non-collider on U is in $\mathcal{S}$. 
\end{itemize}
If the above conditions do not hold we call $X$ and $Y$ conditionally 
independent given the set $\mathcal{S}$ (denoted by $\mathrm{I}(X,Y|\mathcal{S})$). 
Marginal dependency between two nodes can be encoded 
by any path U with no V-structures.
In Figure~\ref{fig:dependencies}b six conditional (in)dependence statements are highlighted in a simple DAG. In the caption of Figure~\ref{fig:dependencies} one of the statements is proved at the hand of conditions (1) and (2). 
\subsection*{Dependencies in PN}
In a PN two nodes $X$ and $Y$ are conditionally dependent given a set $\mathcal{S}$ (denoted by $\mathrm{D}(X,Y|\mathcal{S})$) if and only if there exists a path $\text{U}^*$ between $X$ and $Y$ satisfying:
No node $Z \in \mathcal{S}$ is on $\text{U}$.
If the above condition do not hold we call $X$ and $Y$ conditionally 
independent given the set $\mathcal{S}$ (denoted by $\mathrm{I}(X,Y|\mathcal{S})$). Marginal dependency between two nodes can be encoded by any path U. In Figure~\ref{fig:dependencies}d four conditional (in)dependence statements are highlighted in a simple pairwise Markov network. 

\subsection*{Learning with Hill Climbing and Glasso}
\begin{algorithm}[!ht]
\caption{Hill Climbing \cite{scutari_who_2019}}
\label{algo:greedy}
\textbf{Input:} a data set $\D$ from $\X$, an initial (usually empty) DAG $\G$ and a score function $\Score(\G, \D)$ as given in equation (\ref{eq:bic}).\\
\textbf{Output:} the DAG $\Gmax$ that maximises $\Score(\G, \D)$.
\begin{enumerate}
  \item Compute the score of $\G$, $S_{\G} = \Score(\G, \D)$, and set $\Smax = S_{\G}$ and $\Gmax = \G$. \label{step:setuphc}
  \item Repeat as long as $\Smax$ increases:
  \label{step:hc}

      \begin{enumerate}
        \item for every (or some; simple hill climbing) possible arc addition, deletion or reversal in $\Gmax$ resulting in a DAG:

        \begin{enumerate}
            \item compute the score of the modified DAG $\G^*$, $S_{\G^*} = \Score(\G^*, \D)$:
            \item if $S_{\G^*} > \Smax$ and $S_{\G^*} > S_{\G}$, set $\G = \G^*$ and $S_{\G} = S_{\G^*}$.
        \end{enumerate}
        \item if $S_\G > S_{\mathit{max}}$, set $\Smax = S_\G$ and $\Gmax = \G$.
      \end{enumerate}

\end{enumerate}
\end{algorithm}

\begin{algorithm}[!ht]
\caption{Graphical Lasso}
\label{algo:glasso}
\textbf{Input:} The sample correlation matrix $\S$ generated from a data set $\D$ from $\X$ and the penalization coefficient $\lambda$. \\
\textbf{Output:} The estimated precision matrix $\bm{\Theta}$ (in binary format the undirected PN graph) that maximises the $\Score(\bm{\Theta}, \D)$ as given in equation (\ref{eq:score_lasso}).
\begin{enumerate}
  \item Start with $\bm{W} = \S + \lambda \bm{I}$. The diagonal of $\bm{W}$ remains unchanged in what follows. \label{step:setupgl}
  \item Repeat until convergence:
  \label{step:gl}

    For each $j= 1,2,\dots,N,1,2,\dots,N,\dots$:
      \begin{enumerate}
        \item Reorganize the matrix $\bm{W}$ in $\bm{W}_{11}$ (all but the jth row and column of $\bm{W}$), and $\bm{w}_{12}$ and $\bm{w}_{21}$, the jth row and column without the diagional element $w_{22}$. Do the same  for $\S$.
        \item Solve the lasso regression problem $\min_{\bm{\beta}}\{\frac{1}{2}\|\bm{W}_{11}^{1/2}\bm{\beta}-\bm{b}\|^2+\lambda\|\bm{\beta}\|_1\}$ where $\bm{b}=\bm{W}_{11}^{-1/2}\bm{s}_{12}$,
        this gives a $(N-1)$ -vector solution $\bm{\beta}$:
        \item Fill in the corresponding row and column of $\bm{W}$ using $\bm{w}_{12}=\bm{\beta} \bm{W}_{11}$.
      \end{enumerate}
    \item Finally, using the notation of step \ref{step:gl}(a) for $\bm{\Theta}$, for each $j$, first recover $\theta_{22}$ from the equation $1/\theta_{22} = w_{22} - \bm{w}_{12}\top\bm{\beta}$ and then recover $\bm{\theta}_{12}$ from $\bm{\theta}_{12} = -\bm{\beta}\theta_{22}$.

\end{enumerate}
\end{algorithm}

At the hand of Algorithm \ref{algo:greedy} and \ref{algo:glasso} we outline Hill Climbing and Graphical Lasso. For a more detailed description --and explanation of the equalities in Glasso-- we respectively refer the reader to \cite{russell_artificial_1995} and \cite{friedman_sparse_2008}. The input of both algorithms consists of the dataset  $\cal D$ (the sample correlation matrix $\S$ in Algorithm  \ref{algo:glasso} is just $(1/(n-1)){{\cal D}^\top} \cal D$ for standardized variables) consisting of $n$ independent samples of the multi Gaussian variable $\X$
and a score function to optimize. 
The output of HC is a DAG, whereas the output of Glasso is the estimated precision matrix $\bm{\Theta}$, which, in binary format, is the adjacency matrix of the associated undirected graph. 

Hill Climbing simply visits all (or some; `simple' Hill Climbing) neighbouring networks that differ on one edge of the current network (step 1) and moves then to the network with highest score -- or directly to the first network found with better score in the case of simple HC (step 2). The algorithm stops when no neighbouring network has higher score than the current network. This could be at a local optimum. 

Glasso transforms the initial score function (equation (\ref{eq:score_lasso})) to a lasso problem and applies a coordinate descent approach to solve the problem: the algorithm fixes all dependencies in the current estimate of the correlation matrix $\bm{W}$ except those of one variable (coordinate), i.e. except one column and row (step 2a). Then it estimates the dependencies of this variable that best solves the element wise lasso regression problem (step 2b) and fills in the corresponding row and column in the updated correlation matrix $\bm{W}$ (step 2c).
Next, it moves to the next coordinate and solves the same problem, this time with the former solution integrated in the fixed coordinates (integrated in $\bm{W}_{11}$). This process (step 2) is repeated until convergence. Finally, in the last cycle, the row $\bm{\theta}_{12}$ and diagonal element $\theta_{22}$ in $\bm{\Theta}$ are recovered from $\bm{W}$ and $\bm{\beta}$ (step 3). 

\subsection*{Transformation of probabilistic BN model to probabilistic PN model}
Moralization turns the graph of a directed Gaussian Bayesian network into the graph of an undirected Markov network. Moralization yields the introduction of an undirected edge between any two nodes with a common child and subsequently ignorance of edge directions. Thus, each set of parents and childs $(\XPi)$ is a fully connected subset in the moral graph.  The moral graph M(BN) of a BN is a minimal I-map, however the mapping is not necessarily perfect; not all independencies in the BN are necessarily covered in M(BN). 

An undirected PNM can be asociated with the moral graph in more than one way. 
To asociate the M(BN) with the special case of a probabilistic PN model that encodes the JPD formulated in equation (\ref{eq:mvg-prec}), i.e. a \textit{pairwise} Markov network, the parameterset $(\beta,\nu)$ of the initial BN has to be transformed. The following equality between the precision matrix $\bm{\Theta}$ and the parameters $(\bm{\beta},\bm{\nu})$ of a Gaussian Bayesian Network holds\cite{aragam_concave_2015}:
\begin{equation}
\bm{\Theta} = \bm{\Theta}(\bm{\beta},\bm{\nu}) = (\bm{I}-\bm{B})\bm{\nu}^{-1}(\bm{I}-\bm{B})^\top.
\end{equation}
The new weights of the edges and parameters of the \textit{pairwise} Markov Network are the entries of the precision matrix:
\begin{equation}\theta_{ij} = \theta_{ji} = -\frac{\beta_{ij}(1-\beta_{jj})}{\nu_{j}}-\frac{\beta_{ji}(1-\beta_{ii})}{\nu_{i}} + \sum_{k \neq i,j}\frac{\beta_{ik}\beta_{jk}}{\nu_k}\label{eq:prec_entries}\end{equation}
The entry $\theta_{ij}$ is zero if there is no edge ${i,j}$ in M(BN) (Occasionally,  $\theta_{ij}$ can take the value of zero as a result of the matrix summation at the right hand side of equation (\ref{eq:prec_entries})).

In this work we moralize and extract the precision matrix of all BNs that were learned with the Hill Climbing algorithm. In practice we use the R-packages {\sc bnlearn} for the process of moralization and topological analisis and {\sc sparseBNutils} \cite{aragam_learning_2019} for the extraction of the precision matrix.

\subsection*{Log-likelihood definition and calculation}
The likelihood of the data $\D$, given a model $\M$ is the density of the 
data under the given model $\M$: $\Prob(\D\given\M)$. For discrete density 
functions the likelihood of the data equals the probability of the data 
under the model. The likelihood is almost always simplified by taking the natural 
logarithm; continuous likelihood values are typically small and 
differentiation of the likelihood function (with the purpose of a 
maximum likelihood search) is often hard. Log-likelihood values can be 
interpreted equally when the expression is used for model comparison 
and maximum likelihood search as the natural logarithm is a 
monotonically increasing function. 

In the following we explain the calculation of the log-likelihood 
$\LL(\D|\M) = \log P(\D|\M)$ for a PNM ($\M = \mathrm{PNM}$) for a dataset $\D$ 
formed by $n$ independent data realizations 
$\D_k$, $k \in \{1, \dots, n\}$, of the 
$N$-dimensional random vector $\X$, with $\D_k = \{d^k_1\dots d^k_N\}$ 
and $d^k_i$ the $k$-th realization of variable $X_i \in \X$. We have

\begin{eqnarray}
\log \Prob(\D\given \mathrm{PNM}) &=& \log \Prob(\D_1, \dots, \D_n\given \mathrm{PNM}) = 
\log \prod_{k=1}^{n}\Prob(\D_k\given \mathrm{PNM}) \nonumber \\ 
&=&  \sum_{k=1}^{n}\log\Prob(\D_k\given \mathrm{PNM}) = \sum_{k=1}^{n}\log\Prob_{\mathrm{PNM}}(\D_k) 
\end{eqnarray}

with $\Prob_{\mathrm{PNM}}$ the probability density function as modelled 
by the corresponding PNM
with a Gaussian multivariate probability. 
In this work we considered two types of PNMs, precision and Bayesian 
PNMs, deduced from PNs and BNs graphs, respectively. 
In the case of a PGNM given by a PN  we get:

\begin{eqnarray}
\LL_\mathrm{PN}(\D\given \mathrm{PNM}_\mathrm{PN}) & = &
\sum_{k=1}^{n}\log\Prob(\D_k\given \mathrm{PNM}_\mathrm{PN}) \nonumber \\
& = &\sum_{k=1}^{n}\log\{(2\pi)^{-N/2} \det(\bm{\Theta})^{1/2} \nonumber\\
& & \times \exp \{-1/2\sum_{i}^{N}\theta_{ii}(d^k_i)^2-\sum_{i<j}\theta_{ij}d^k_id^k_j\}\}.
\label{eq:loglik_pn}
\end{eqnarray}

Entries in the sum are evaluations of the multivariate normal density function and executed with the R-package {\sc mvtnorm} \cite{computation_genz_2009}.

In the case of a PGNM given by a BN, from equation~(\ref{eq:mvg-fac}), we have
\begin{eqnarray}
\LL_{\mathrm{BN}}(\D\given \mathrm{PNM}_{\mathrm{BN}}) &=& \sum_{k=1}^{n}\log\Prob(\D_k\given \mathrm{PNM}_{\mathrm{BN}}) \nonumber \\&=&
\sum_{k=1}^{n}\log\prod_{i=1}^{N} \Prob_i(X_i = d^k_i \given \Pi_{X_i} = 
d^k_{\Pi_{X_i}}) \nonumber \\
&=& \sum_{k=1}^{n}\sum_{i=1}^{N} \log\Prob_i(X_i = d^k_i \given \Pi_{X_i} = 
d^k_{\Pi_{X_i}}),  
\label{eq:loglik_bay}
\end{eqnarray}
where $d^k_{\Pi_{X_i}}$ is a subset of $\D_k$ containing the $k$-th data 
realization of the parentset $\Pi_{X_i}$ of $X_i$. From equation~(\ref{eq:mvg-loc}) 
we know that the conditional univariate densities in the sum in equation~(\ref{eq:loglik_bay}) 
are univariate normal and we execute them with the basic R-package {\sc stats}.

\bibliography{bibliography}



\section*{Acknowledgements}

CEG would like to acknowledge the support of the funding from the Spanish Agencia Estatal de Investigaci\'on through the Unidad de Excelencia Mar\'ia de Maeztu with reference MDM-2017-0765.

\section*{Author contributions statement}

CEG and JMG conceived and designed the study. CEG wrote the main manuscript text. JMG revised the manuscript. All authors read and approved the final manuscript.

\section*{Additional information}

\textbf{Competing interests} The authors declare that they have no competing interests. \textbf{Correspondence} Correspondence and requests for materials
should be addressed to CEG.~(email: catharina.graafland@unican.es).





\end{document}